\documentclass[11pt, oneside]{article}   	
\usepackage{geometry}                		
\usepackage{graphicx}				
					          
\usepackage{amssymb}

\title{The three-fold theoretical basis of the Gravity Probe B gyro precession calculation}
\author{Ronald J. Adler* \\
Hansen Experimental Physics Laboratory, Gravity Probe B Mission, \\ Stanford University, Stanford, California 94309, \\ and
\\Department of Physics and Astronomy, \\ San Francisco State University, San Francisco, California, \\and\\ Kavli Institute for Particle Astrophysics and Cosmology,\\
Stanford University, Stanford California 94035
94132}
\date{ May 21, 2014}				
\begin{document}
\maketitle
\begin{abstract}
The Gravity Probe B (GP-B) experiment is complete and the results are in agreement with the predictions of general relativity (GR) for both the geodetic precession, 6.6 arcsec/yr to about 0.3\%, and the Lense-Thirring precession, 39 marcsec to about 19\%. This note is concerned with the theoretical basis for the predictions. The predictions depend on three elements of gravity theory, firstly that macroscopic gravity is described by a metric theory such as general relativity, secondly that the Lense-Thirring metric provides an approximate description of the gravitational field of the spinning earth, and thirdly that the spin axis of a gyroscope is parallel displaced in spacetime, which gives its equation of motion. We look at each of these three elements to show how each is solidly based on previous experiments and well-tested theory. The agreement of GP-B with theory strengthens our belief that all three elements are correct and increases our confidence in applying GR to astrophysical phenomena. Conversely, if GP-B had not verified the predictions a major theoretical quandary would have occurred.
\end{abstract}

*electronic mail address: adler@relgyro.stanford.edu or gyroron@gmail.com 

\section{Introduction}

After 47 years the Gravity Probe B (GP-B) experiment is complete.\cite{1,2} The data analysis was more demanding than expected, due largely to complicating classical effects, for example electric charge on the rotors and housing as discussed at length in other papers in this volume.\cite{3,4} The bottom line is that the predictions of general relativity (GR) for the geodetic effect are confirmed to about 0.3\% and for the Lense-Thirring (LT) effect to about 19\%. In this paper we will be concerned with what that experimental confirmation implies for gravity theory in general and in particular for GR.\cite{5,6,7} Our aim in this paper is to focus on how the prediction of the gyro precessions come about and what assumptions are needed, and thus to what extent the experiment verifies theory, in particular GR. 

Three key elements enter the calculation of the precession. The first is the most fundamental, that macroscopic gravity is described by a geometric theory, and specifically a metric theory.\cite{8,9,10,11,12} The second key element is that the specific metric for a nearly spherical spinning body, such as the earth, is the approximate one found in 1918 by Lense and Thirring using linearized GR.\cite{13,14,15,16} The third key element is that the spin vector of a gyro is parallel displaced in spacetime, which implies that the equation of motion for the spin is that its covariant derivative is zero.\cite{17,18} We will focus on analyzing how well founded are these elements. Our discussion will not be exhaustive since the literature contains many variations on the theme. Thus, unfortunately, we cannot reference many interesting and important theoretical papers on the subject. Just a few are listed in the references.\cite{19,20,21,22,23,24,25}

There are of course many small corrections to the precession calculation, due for example to the multipole moments of the real earth, rather than the idealized spherical earth, due to the presence of the sun and moon, etc.\cite{14,15,26} There are also small corrections to the geodesic motion of spinning test bodies that are relevant to the equivalence principle, which we discuss in sec.5. 

Throughout this paper we will make use of appropriate approximations to gravity theory since the field of the earth is quite weak, and we will also make use of the fact that the earth and the gyro move at low velocity. As in most theory papers we will use units in which $c=1$. 

\section{Metric theory in general}

It has been standard lore since the formulation of GR that gravity is described by a metric theory.\cite{8,9,10} The most obvious motivation for this assumption is the so-called weak equivalence principle (EP), or more accurately the ``universality of free fall"  for test bodies in a gravitational field. A metric theory provides an obvious elegant explanation for why the trajectories of test bodies in a gravitational field are independent of their masses and also various internal properties. The EP has been tested to impressive accuracy, better than about $10^{-12}$.\cite{5} This may be improved to   $10^{-15}$ in an upcoming free-fall satellite experiment, perhaps to $10^{-15}$ by future atomic beam interferometry, and hopefully to $10^{-18}$ in a more accurate satellite experiment in the more distant future.\cite{27,28,29,30} That $10^{-18}$ estimate seems to be the present anticipated limit.

Various authors, notably Jordan and later Brans and Dicke, have suggested that a scalar field should be added to the description of gravity.\cite{31,32,33,34} Some authors are of the opinion that string theory motivates such a modification, but there is as yet no experimental evidence to support string theory and no experimental evidence for a scalar field in gravity theory.\cite{35}  Will discusses both scalar tensor theory and its experimental tests from the PPN perspective.\cite{36} In summary, so far the evidence is that a pure metric theory is adequate to describe macroscopic gravity, but the question remains interesting and open to experiment. 

\section{Experimental status of GR and the Schwarzschild metric in a nutshell}

This section will be a shamelessly short and over-simplified summary of parts of the book and arxiv paper by Will, leading to the conclusion that the Schwarzschild metric of GR has been quite well tested by observation and experiment.\cite{5} Unfortunately all of the evidence involves weak fields and rather low velocities, and there are as yet no precision tests of strong gravity; observations of black holes may lead to such tests in the future by studying, for example, the motion of material near the surfaces of black holes.\cite{37,5}

The Òclassical testsÓ of GR, the gravitational red shift, the orbit of Mercury and the deflection of light by the sun, are all based on the Schwarzschild metric, obtained in 1916, which describes the metric field of a spherically symmetric non-spinning body.\cite{38,5,12} In the standard coordinates the metric is
\begin{equation}
\label{1}
ds^2=(1-2m/r_s)dt^2-(1-2m/r_s)^{-1}dr_s^2-r^2_sd\theta ^2 -r_s^2\sin^2\theta d\varphi ^2 ,
\end{equation}
 where $m$ is termed the geometric mass; $M$ is the mass of the body and $G$ is NewtonÕs constant. In the so-called isotropic coordinates, which are convenient for comparison with observation, the metric is,\cite{39} 
\begin{eqnarray*}
\label{2}
ds^2= \frac{(1-m/2r)^2} {(1+m/2r)^2}-(1+m/2r)^4d\vec{r}^2 
\end{eqnarray*}
\begin{equation}
\label{2a}
=\left( 1-\frac{2m}{r}+\frac{2m^2}{r^2}+ ...\right)dt^2-\left( 1+\frac{2m}{r}+ \frac{3m^2}{2r^2}...\right)d\vec{r}^2 .
\end{equation}
The power series expansion in the last line is useful and valid for distances far from the central body where $m/r<<1$. Eddington re-expressed (2) in terms of 3 dimensionless parameters, $\alpha,\beta,\gamma$, as \cite{40}
\begin{equation}
\label{3}
ds^2=\left( 1-\alpha\frac{2m}{r}+\beta\frac{2m^2}{r^2}+ ...\right)dt^2-\left( 1+\gamma\frac{2m}{r}+ ...\right)d\vec{r}^2 .
\end{equation}
The parameter $\alpha$  is a measure of the distortion of time due to gravity, but the way in which it enters the metric makes it impossible to separate from NewtonÕs constant $G$, and as a result it may be taken to be 1; we will retain it only as a bookkeeping device, as we will discuss below. The parameter $\beta$ is a measure of the nonlinearity of time distortion effects; $\gamma$  is a measure of the distortion of space to first order. In GR all the parameters are equal to unity, $\alpha=\beta=\gamma$. The quadratic term in the spatial part of the metric (2) is not yet measurable and does not appear in (3), nor do any other higher order terms. 

The Eddington form (3) of the metric can be viewed in two ways. The first is as a bookkeeping device to see how various physical predictions depend on properties of GR; for example the precession of the orbit of mercury depends on the combination $\beta+\gamma$ so we may say that the nonlinearity of time distortion and linear space distortion are being tested. The GPB experiment measured $\gamma$  as we will discuss below. 

The second point of view of (3) is that the parameterization could describe a metric theory other than GR, and is thus more general. The parameterized post Newtonian (PPN) theory of Nordtvedt, Will and others carries this viewpoint to a high level of generality and sophistication with the use of about 9 parameters that can be tested experimentally.\cite{5,6,7} Moreover the PPN approach involves an expansion in powers of  $m/r$ and $1/c$ and often provides clear intuitive understanding of physical effects, analogous to Newtonian theory.\cite{41} 

As we have noted, previous observations and experiments in the solar system and observations of pulsar systems are in agreement with GR, but all involve weak fields, even the pulsar systems. As a measure of the accuracy of such tests the Eddington parameters $\beta,\gamma$, which are predicted to be 1 by GR, are found from various observations to be   
$|\gamma-1|< 2.3\times10^{-5}$ and $|\beta-1| < 8\times10^{-5}$.\cite{5} Thus the approximate Schwarzschild metric (3) is well verified, and the Eddington parameters are very close to 1. 

\section{The Lense-Thirring metric from several points of view}	

The Schwarzschild metric in (1) and the approximation (3) only describe the metric exterior to a spherically symmetric non-spinning body and therefore allow us to calculate only the geodetic part of the gyro precession. The LT part of the gyro precession depends on a generalization of the Schwarzschild metric due to the spin of the source. For the earth, which is not very massive and spins slowly, the modification is quite small, making its effect on the gyro excedingly difficult to detect. 

The metric for the exterior of a spinning spherical body was first obtained by Lense and Thirring in 1918 using linearized GR.\cite{13} They worked to lowest order in the gravitational fields and velocities and obtained a metric that we may write in spherical coordinates as 
\begin{equation}
\label{4}
ds^2=(1-2m/r)dt^2-(1+2m/r) d\vec{r}^2+2\left(\frac{2GJ}{r}\right)\sin^2\theta d\varphi dt, 
\end{equation}
where $J$ is the angular momentum of the spinning source body. 

A more general version of the LT metric can also be obtained using the so-called gravito-electromagnetic (GEM) approximation, which applies for weak fields and slowly moving bodies, and in which many equations are similar to those of classical electrodynamics.\cite{14,15} In the GEM approximation the metric may be written as
\begin{equation}
\label{5}
ds^2=(1+2\phi)dt^2-(1-2\phi) d\vec{r}^2+2(\vec{h} \cdot d\vec{r} )dt .
\end{equation}
Here $\phi$ is the Newtonian potential outside of the body and $\vec{h}$ is called the gravito-magnetic 3-vector potential, analogous to the 3-vector potential of electrodynamics; $\phi$ and $\vec{h}$  may be defined as 
\begin{equation}
\label{6}
\phi(\vec{r})=-G\int \frac{\rho(\vec{r'})d^3 r'}{|\vec{r} - \vec{r'}|} , \; 
\vec{h}(\vec{r})=4G\int \frac{\rho(\vec{r'})\vec{v}(\vec{r}')d^3 r'}{|\vec{r} - \vec{r'}|} . 
\end{equation}
  
There is one approach to the LT metric that we believe is worth further discussion because it rests on a solid semi-empirical basis and is thus nearly independent of theory. \cite{14,15} This approach depends on three well founded assumptions that are motivated by experiment and established theory. It also assumes the weak fields and low velocities appropriate for GP-B, and makes clear why the LT metric depends (to an excellent approximation) only on the Eddington parameters $\alpha =1$ and $\gamma$ and not on any independent new parameter related to the spin and gravito-magnetism.  First, the metric for the exterior of a small spherical body, essentially a point mass, is given by the approximate expression (3) for the Schwarzschild geometry, which is well verified by experiment as we have discussed. Second, the metric for such a body in motion is given by a Lorentz transformation of (3) in accord with basic relativity theory. Third, due to the weakness the fields, the metric for many such small bodies or point masses is a superposition of the individual metric for each body, analogous to the superposition of potentials in Newtonian theory. 

We will review the logic of the derivation in some detail. We begin with the Eddington form (3) for a point mass at rest and apply a Lorentz transformation in the $x$ direction, to first order in velocity $v$, 
\begin{equation}
\label{7}
t_s=t-vx , \; x_s=x-vt , 
\end{equation}							 
and obtain the metric for a slowly moving point mass, to lowest order in $m/r$  and velocity $v$,
\begin{equation}
\label{8}
ds^2=\left(1-\alpha\frac{2m}{r}\right)dt^2-\left(1+\gamma\frac{2m}{r}\right)d\vec{r}^2+(\alpha + \gamma)\left(\frac{4m}{r}\right)(\vec{v}\cdot d\vec{r})dt. 
\end{equation}
The generalization to motion in any direction $\vec v$ is obvious from (8). Since the fields are assumed to be weak we superpose the fields of many such point masses just as in Newtonian theory, using the recipes
\begin{equation}
\label{9}
\frac{-GM}{r} \rightarrow-G\int \frac{\rho(\vec{r'})d^3 r'}{|\vec{r} - \vec{r'}|}=\phi(\vec{r}) , \; 
\frac{4GM\vec{v}}{r} \rightarrow4G\int \frac{\rho(\vec{r'})\vec{v}(\vec{r}')d^3 r'}{|\vec{r} - \vec{r'}|} =\vec{h}(\vec{r}) . 
\end{equation}
These are exactly the same functions that occur in (6) so we obtain
\begin{equation}
\label{10}
ds^2=(1+2\alpha \phi)dt^2-(1-2\gamma \phi) d\vec{r}^2+(\alpha + \gamma)(\vec{h} \cdot d\vec{r} )dt ,
\end{equation}
which is the same as (5) but includes the Eddington parameters. Our derivation produced no new parameters in the last expression, and the effects of gravito-magnetism are parameterized by $\alpha+\gamma=1+\gamma$. 

Within the broader context of the PPN formalism there is another parameter that could be included in the above discussion, called $\alpha_1$, which is related to the possible existence of a preferred inertial reference frame. It would entail adding $\alpha_1/4$ to $\gamma$  in (8), but observations constrain $\alpha_1$ to be less than $10^{-4}$  so it is not relevant to GP-B and we will not include it here.\cite{42,43,6} 

We note that the expression (5) is obviously not limited to a spherical body; it presumes only weak fields and low velocities. From (5) the multipole corrections for a slightly non-spherical body such as the earth have also been worked out.\cite{15}

For a nearly spherical body such as the earth the metric (10) may also be conveniently written in spherical coordinates, again to lowest order in $m/r$, as 
\begin{equation}
\label{11}
ds^2=(1-2\alpha m/r)dt^2-(1+2 \gamma m/r) d\vec{r}^2+(\alpha+\gamma)\left(\frac{2GJ}{r}\right)\sin^2\theta d\varphi dt, 
\end{equation}
which is the LT form of the metric (4), but with Eddington parameters included. 

The LT metric has been obtained in various other ways. For example it can be derived by an expansion from the exact Kerr metric for a spinning black hole.\cite{16}

\section{The equation of motion for the spin} 
					
The precession of a gyro in GR and other metric theories is an extraordinary effect: in Newtonian theory there is no analog. For example a gyro in a uniform Newtonian force field does not precess. This is one of the reasons that the GPB experiment is particularly interesting to theorists. Moreover the precession is a so-called Machian effect: the presence of the rotating earth has an effect on determining the local inertial frame, in sharp contrast to Newton's absolute space. In the context of GR and similar theories the behavior of the gyro spin has been studied in a number of ways, but we will focus on only two of them and briefly mention a third.\cite{17,18,19,20,21,22,23,24,25,44,42} The conclusion is that the spin four-vector $S^{\mu}$ is parallel displaced along its trajectory in spacetime, which gives a simple equation for the gyro precession. This conclusion is independent of theoretical details and not limited to GR. 

Our first argument is based on simplicity and general covariance, and makes the equation of motion intuitively obvious.  Let us first consider a general affine space, in which there is a law of parallel displacement using coefficients of affine connection. The space need not even have a metric. In such a space there is only one privileged or special curve, a geodesic; the geodesic may be defined as that curve for which the tangent vector (or four-velocity) $u^{\mu}=dx^{\mu}/ds=\dot{x}^{\mu}$ is displaced parallel to itself along the curve, or that the curve is Òparallel to itself.Ó This implies that the covariant derivative of $u^{\mu}$  along the curve is zero, or 
\begin{equation}
\label{12}
Du^{\mu}/Ds=du^{\mu}/ds+\Gamma^{\mu}_{\omega \sigma} u^{\omega}u^{\sigma}=0 \; \; or \; \; 
\ddot{x}+\Gamma^{\mu}_{\omega \sigma} \dot{x}^{\omega} \dot{x}^{\sigma}=0.
\end{equation}
In the most general case the arc length may be replaced by any invariant parameter.\cite{45}

We can use similar reasoning to heuristically motivate an equation for the gyro spin $S^{\mu}$.
First we note that in the rest frame of the gyro the four-velocity and spin vector are 
\begin{equation}
\label{13}
u^{\mu}=(1,0,0,0) \; , \; S^{\mu}=(0,\vec{S}) \; , \;  S^{\mu}u_{\mu}=0 , \; (gyro\; rest\; frame) \; ,
\end{equation}
and since $S^{\mu}u_{\mu}=0$ is a covariant expression it holds in any frame. It is a well-known property of parallel displacement that that if two vectors are parallel displaced together then their inner product does not change; it thus becomes natural to demand that the spin $S^{\mu}$  be parallel displaced along the geodesic path of the gyro, along with the four-vector velocity. Then the spin equation of motion and the orthogonality condition are
\begin{equation}
\label{14}
DS^{\mu}/Ds=dS^{\mu}/ds+\Gamma^{\mu}_{\omega \sigma} u^{\omega}S^{\sigma}=0 \; , S^{\mu}u_{\mu}=0\; .
\end{equation}
We stress that parallel displacement is a sufficient but not a necessary condition that the inner product $S^{\mu}u_{\mu}$ remains zero along the trajectory. Note also that (14) is a rather general result and does not depend on any particular theory of gravity, but of course its application to a particular problem will use affine connections which do depend on the specific theory. 

We can also phrase the argument in terms of general principles. In sec. 2 we mentioned the weak equivalence principle (EP) or universality of free fall, which states that the trajectories of test bodies in a gravitational field are independent of their masses and various internal properties. Thus in a freely falling lab or reference frame test bodies behave as if there were no gravitational field present. The phenomenon has become familiar in television broadcasts from orbiting spacecraft. Conversely, in an accelerated lab or reference frame test bodies behave as if there were a gravitational field present. Einstein proposed an extended version of the equivalence principle, called the Einstein equivalence principle (EEP) that assumes ``complete physical equivalence of a gravitational field and a corresponding acceleration of the reference system." The EEP includes nongravitational phenomena, such as electromagnetism, as well as gravitational phenomena. It leads to a Òprinciple of general covarianceÓ that has proven to be very powerful in formulating nongravitational physical laws, such as Maxwellian electrodynamics in the presence of a gravitational field. To use the principle of general covariance one writes an equation that is known to be correct in the absence of gravity, and takes it to be true for a freely falling reference frame in which there are no effects of gravity, according to the EEP. Then if the equation is expressed in generally covariant form it must also be correct in any reference frame.\cite{17,12}

It is easy to apply the ideas of the EEP and general covariance to the equation of motion of the gyro and to its spin vector, as elucidated clearly by Weinberg.\cite{17} In a space with no gravitational field it is obvious that both the four-vector velocity $u^{\mu}$ of the gyro and the spin vector $S^{\mu}$ should be constant, or $du^{\mu}/ds=0$ and $dS^{\mu} / ds=0$. According to the principle of general covariance the correct generalization of these equations in a gravitational field is obtained simply by replacing the Lorentz metric by the general Riemannian metric and the ordinary derivative by a covariant derivative, making the equations generally covariant. The result is (12) and (14). 

Our second argument was given by Papapetrou and is much more physical.\cite{18} His argument does not depend explicitly on the field equations of GR, but on the conservation of energy momentum, expressed as the tensor equation 
$(T^{\mu \nu})_{;\nu}=0$; the conservation equation does of course follow from the field equations of GR. Papapetrou analyzed a small ball of material, making few assumptions about its internal structure, and derived the correct geodesic equation of motion as a first approximation by ignoring various three-space moments of the ball. (He called it the monopole approximation.) He then took account of internal structure and motion of the material in the ball to lowest order in the size of the ball, including first moments in three-space, and thereby obtained an equation of motion for its second rank anti-symmetric spin tensor $\tilde{S}^{\mu \nu}$; that tensor is defined as 
\begin{equation}
\label{15}
\tilde{S}^{\mu \nu}=\int dV(\delta x^{\mu} u^{\nu}-\delta x^{\nu} u^{\mu})\rho \;,
\end{equation}
where the integral is over the three-space volume of the ball, $\delta x^{\mu}$ is the position in the ball relative to its center of mass, $u^{\nu}$ is the four-velocity of the ball material, and $\rho$  is its density. The equation of motion for the spin tensor that he obtained is 
\begin{equation}
\label{16}
D\tilde{S}^{\mu \nu}/Ds+u_{\rho} (u^{\nu} D\tilde{S}^{\rho \mu}/Ds-u^{\mu} D\tilde{S}^{\rho \nu}/Ds)=0 \; .
\end{equation}

We still need to relate the spin vector $S^{\mu}$ to the spin tensor $\tilde{S}^{\mu \nu}$ and also relate the equations that they obey, that is (14) and (16). The spin vector has three independent components and the tensor has six, but only three of them determine the angular position of the gyro. We can find the desired relation by using the low velocity limit as a guide. The tensor $\tilde{S}^{\mu \nu}$ is antisymmetric, the time displacement on a spatial surface is $\delta x^0=0$, and $u^0 \approx c =1$. Hence $\tilde{S}^{\mu \nu}$ is approximately
\begin{equation}
\label{17}
\tilde{S}^{00}=0 \;,  \tilde{S}^{0j}=\int dV \rho \delta x^j = 0\; , 
\tilde{S}^{ij}=\int dV \rho( \delta x^i v^j - \delta x^j v^i) .
\end{equation}
The second relation in (17) follows since $\delta x^j$ is measured from the center of mass of the body. Thus the spatial part of the spin tensor is the familiar angular momentum tensor of three-dimensional mechanics. The three-vector angular momentum is related to it by the well-known equation
\begin{equation}
\label{18}
S^i=(1/2)\epsilon_{ijk} \tilde{S}^{jk} .
\end{equation}
What we now need is a covariant generalization of (18) to relate $S^{\mu}$ to $\tilde{S}^{\mu \nu}$. A moment's thought provides an answer, which is 
\begin{equation}
\label{19}
S^{\mu}=(-1/2)u_\rho {e^{\rho \mu}}_{\alpha \beta} \tilde{S}^{\alpha \beta} , \; 
e_{\alpha \beta \gamma \delta}\equiv \sqrt{-g} \epsilon_{\alpha \beta \gamma \delta} , 
\end{equation}
where $\epsilon_{\alpha \beta \gamma \delta}$ is the usual Levi-Cevita alternating symbol and $e_{\alpha \beta \gamma \delta}$ is the Levi-Cevita tensor. \cite{24,46} Equation (19) is clearly a generally covariant expression. Moreover it is obvious from the antisymmetry of the Levi-Cevita tensor that
\begin{equation}
\label{20}
S^{\mu}u_{\mu}=0 ,
\end{equation}
which is the same as the orthogonality relation in (14). 

Finally we can obtain the parallel displacement relation (14) from Papapetrou's equation (16). To do this we first note that the Levi-Cevita tensor has a zero covariant derivative, as does the four-velocity vector $u^{\mu}$ along a geodesic.\cite{46} Then from (16) the covariant derivative of the vector $S^{\mu}$ is
\begin{eqnarray*}
\label{21}
DS^{\mu}/Ds=(-1/2)u_\rho {e^{\rho \mu}}_{\alpha \beta} D\tilde{S}^{\alpha \beta}/Ds 
\end{eqnarray*}
\begin{equation}
\label{21a}
=(-1/2)u_\rho {e^{\rho \mu}}_{\alpha \beta} (u_\sigma u^\beta D\tilde{S}^{\alpha \sigma}/Ds- 
u_\sigma u^\alpha D\tilde{S}^{\beta \sigma}/Ds)=0 .
\end{equation} 
Thus the Papapetrou analysis leads to the same equation we obtained previously; the parallel displaced spin vector equation (14) obtained from general principles also follows from a more detailed ``nuts and bolts" analysis

The spin equation (14) implies an important fact about the gyro precession since it is homogeneous in the spin $S^\mu$. The vector $S^\mu$ and tensor $\tilde{S}^{\mu \nu}$ clearly depend on the rotation rate of the gyro, which is clear from the definition in (15). But since the spin equation is homogeneous the angular precession is independent of the magnitude of $S^\mu$, so the gyro spin velocity is, in principle, irrelevant and has no effect on the precession; $S^\mu$  merely serves to define a direction in space. Of course in the real world of experiments the spin velocity may be very important in the accurate measurement of the precession.

Papapetrou noted another fact of interest, that a spinning body does not follow a geodesic exactly, as in (12), but deviates a little due to the interaction between spin, orbital angular momentum and curvature. His equation giving the modified geodesic is the following
\begin{equation}
\label{22}
\frac{D}{Ds}(mu^\alpha +  u_\beta  \frac{D\tilde S^{\alpha \beta}}{Ds}) 
+ \frac{1}{2}  \tilde S^{\mu \nu} u^\sigma {R^\alpha}_{\nu \sigma \mu} =0 .
\end{equation}
As might be expected the extra terms in (22) are far too small to be relevant for the GP-B experiment, or any solar system experiment envisioned at present. In reference \cite{24} Will notes that there is some disagreement about the result (22), and gives further references; he also estimates such effects to be well below $10^{-20} g$. However it is clear that in principle the motion of a body depends on its spin and internal structure, so the EP or universality of free fall cannot be an exact principle but only an extraordinarily accurate approximation. That is, GR transcends the EP; to quote Nordtvedt ``Principles in physics are for when you have no theory." 

Furthermore spin effects such as displayed in (22) may be large for some astronomical systems, such as black holes or neutron stars in close orbit. The gravitational radiation emitted by such bodies during their final inspiral may allow observation of the spin effects, as indicated by numerical GR simulations.\cite{5}

We mention in passing one other interesting approach to the LT gyro spin theory. Murphy, Nordtvedt and Turyshev have used a PPN approach to derive the LT gyro precession, in agreement with the one we give here. The virtue of their quasi-Newtonian derivation is that it shows how the LT gyro precession results from the gravito-magnetic acceleration of each moving point mass in the rotating gyro.\cite{42} They also include the $\alpha_1$  parameter as mentioned previously, which is known to be small from previous observations and has a negligible contribution. 

One extension of GR theory, that in principle could affect the spin equation, involves the concept of torsion; in GR the affine connections are symmetric in the lower indices, but if they are allowed to have an antisymmetric part the result is a more general theory than GR, called Einstein-Cartan theory, which involves the concept of torsion.\cite{47,48}  Many theorists believe torsion should be included in gravity theory, for example to accommodate the spin of particles, although no experiments indicate such a need.\cite{49} Moreover other authors have developed the quantum theory of spin 1/2 particles interacting with gravity without the use of torsion, so torsion appears to be neither observed nor needed for theoretical consistency.\cite{50,51} Of course that does not prove it does not exist in nature. 

In summary, the general spin equation (14) appears to be well founded on both mathematical and physical grounds.

\section{Solving the spin equation for GP-B}					

From the LT metric in (10) and the general spin equation (14) it is straightforward although slightly tedious to calculate the precession of the GP-B gyro in its polar orbit. The gravitational field of the earth is weak so that the expansion of the metric to order $m/r$  is adequate, and the rotational velocity of the earth and the orbital velocity of the satellite are small, so we need only work to first order in $v$. Also we will assume a perfectly circular polar orbit with the gyro spin in the orbital plane.  See fig.1a and also fig.1 of the overview paper by Everitt in this volume. 

We will briefly sketch the calculation following references [15] and [17]. The first step of the calculation is to obtain the affine connections from the LT metric and substitute them into the spin equations (14). This yields the following equation for the space part of the spin vector, written in index notation, 
\begin{equation}
\label{23}
\dot{S}^i=\left[ \gamma V^i (\phi_{,k} S^k ) +  \gamma S^i (\phi_{,k }V^k ) -
(\alpha + \gamma)\phi_{,i} (S^k V^k)\right]
+ (1/4)(\alpha + \gamma)(h_{i,k} - h_{k,i})S^k.
\end{equation}
 Here $V^k$ is the 3-vector velocity of the gyro. In three-vector notation (23) is
\begin{equation}
\label{24}
\dot {\vec S}=\left[ \gamma \vec V (\nabla{\phi} \cdot \vec S ) +  \gamma \vec S (\nabla{\phi} \cdot \vec V) - 
(\alpha + \gamma) \nabla \phi (\vec S \cdot \vec V)\right]
+ (1/4)(\alpha + \gamma)(\nabla \times \vec h ) \times \vec S.
\end{equation}
The terms that contain the Newtonian potential $\phi$  contribute to the geodetic precession, and those that contain the gravito-magnetic potential $\vec h$ contribute to the LT precession.  Next we split the square bracket in (24) containing the potential $\phi$ into two parts, anti-symmetric and symmetric in the pair of vectors $\vec V$ and $\nabla \phi$, and write it as 
\begin{eqnarray*}
\label{25}
(\gamma + \alpha /2)\left[ \vec V (\nabla{\phi} \cdot \vec S)- \nabla  \phi (\vec S \cdot \vec V )\right ] +
\left \{ \gamma \vec S (\nabla \phi \cdot \vec V)  -\alpha /2 [ \vec V (\nabla \phi \cdot \vec S ) + \nabla \phi (\vec S \cdot \vec V) ]  \right \}
\end{eqnarray*}
\begin{equation}
\label{25a}
=(\gamma + \alpha /2)(\nabla \phi \times \vec V ) \times \vec S +
\left \{ \gamma \vec S (\nabla \phi \cdot \vec V)  -\alpha /2 [ \vec V (\nabla \phi \cdot \vec S ) + \nabla \phi (\vec S \cdot \vec V) ]  \right \}
\end{equation}
The curly bracket in (25), which is symmetric in $\vec V$ and $\nabla \phi $, averages to zero over a circular orbit.
More generally, in the Newtonian approximation $\dot{ \vec V}=-\nabla \phi $, and also the change in the spin $\vec S$ is extremely slow; these two facts allow us to express the curly bracket as a time derivative, so it must average to zero over  general orbits. We will henceforth ignore it. Then (24) simplifies to 
\begin{equation}
\label{26}
\dot {\vec S}=\left[ (\gamma + \alpha /2)(\nabla \phi \times \vec V) +  
(1/4)(\alpha + \gamma )\nabla \times \vec h \right]  \times \vec S.
\end{equation}
To make (26) beautiful we define two vector fields, a geodetic vector field a gravito-magnetic vector field, in terms of the Newtonian potential and the gravito-magnetic vector potential, as 
\begin{equation}
\label{27}
\vec \Omega_G =(\gamma + \alpha /2)(\nabla \phi \times \vec V) \; ,
\; \vec \Omega_{LT }=(1/4)(\gamma + \alpha)\nabla \times \vec h .
\end{equation}
Both fields are independent of time for the spherical spinning earth. Then (26) becomes
\begin{equation}
\label{28}
\dot {\vec S}=\left( \vec \Omega_G + \vec \Omega_{LT}\right)  \times \vec S,
\end{equation}
which we recognize as the classical equation for a precessing gyro.  Since that problem is quite well known our problem is nearly solved. 

For the GP-B gyro the precession is extremely slow, so the spin does not change appreciably over the course of many orbits, and we may write the change in $\vec S$ in time $\Delta t$ as 
\begin{equation}
\label{29}
\Delta \vec S=\Delta \vec S_G + \Delta \vec S_{LT}=
(\vec \Omega_G \times \vec S) \Delta t +(\vec \Omega_{LT} \times \vec S) \Delta t ,
\end{equation}
with $\vec S$ treated as a constant. The last expression in (29) defines the geodetic and LT drifts, which are linear in time. 

Consider the geodetic term of (29) first, by far the larger part. For a circular orbit the gravitational force and the velocity are perpendicular, and the geodetic field is thus perpendicular to the orbit plane. The geodetic vector and its magnitude are
\begin{equation}
\label{30}
\vec \Omega_G =(\gamma + \alpha /2)\left (\frac{GM}{r^2} \right) (\hat r \times \vec V) \; ,
\Omega_G =(\gamma + \alpha /2)\left (\frac{GMV}{r^2} \right) .
\end{equation}
The various vector directions are shown in fig.1. The geodetic precession is in the plane of the orbit. 

The LT precession depends on the gravito-magnetic field $\vec \Omega_{LT}$, which varies with position in the orbit. The gravito-magnetic vector potential $\vec h$ of the spinning earth can be calculated in the same way as the vector potential of a spinning ball of charge in electrodynamics. The results for $\vec h$ and $\vec \Omega_{LT}$ are
\begin{equation}
\label{31}
\vec h =\left ( \frac{2G}{r^3} \right )(\vec r \times \vec J )\;, \; 
\vec \Omega_{LT} = (1/2)(\gamma + \alpha)G \left ( \frac{ \vec J}{r^3} - \frac {3 \vec r}{r^5} (\vec r \cdot \vec J) \right ) ,
\end{equation}
where $\vec J$ is the angular momentum of the earth. The gravito-magnetic field has exactly the same shape as a magnetic dipole field, as might be expected. Since the gyro precesses so slowly we need only average $\vec \Omega_{LT}$ over an orbit to obtain the LT precession,  
 \begin{equation}
\label{32}
\left\langle \vec \Omega_{LT}  \right \rangle = (1/2)(\gamma + \alpha) \frac{G \vec J}{2 r^3} ,
\end{equation}and the magnitude of this is the LT precession. The precession is perpendicular to the orbit plane as shown in fig.1. 

Our final results for the precessions are given in (29) to (32) with directions shown in fig.1. It is important to emphasize that for the polar orbit the geodetic and LT precessions are perpendicular; the LT precession is very much smaller than the geodetic and would not be measurable if the two were not accurately perpendicular. 

We will not discuss small and subtle corrections to the basic precessions, such as the effects of the earth multipole moments, the presence of the moon and the sun, variations in the spacecraft altitude and orbital orientation, etc. These are covered in the references and the other papers in this volume.\cite{44,15}
	
\section{Summary and further comments}

This work has focused on the bases for the theoretical predictions of the gyro precession in the GP-B experiment. We have at most only mentioned some of the interesting subtleties and small corrections to the predictions, such as the effect of the sun and the quadrupole moment of the earth, which are covered in the references. We have not found any of the suggestions that the standard results are substantially wrong convincing enough to discuss them; this seems well justified by the experimental results. Some small but possibly interesting and well-founded modifications to the basic predictions might concern a scalar field component added to gravity theory or a torsion related addition to the equation of motion for the spin. The GP-B results indicate that neither of these is presently needed, at least within the accuracy of the experiment. 

\section{Acknowledgements}

The GP-B theory group has provided interesting comments and criticisms on the theory behind the experiment. In particular we thank Robert Wagoner, Paul Worden, and Francis Everitt for patient discussions of the experimental situation, the PPN formalism, and references to the literature. We thank Kenneth Nordtvedt for correspondence and his notes regarding the equivalence principle. In particular his comment that ``Principles in physics are for when you have no theory" is particularly relevant to our discussion in sec.5. Finally, we thank Cliff Will for technical comments and updates on the references.

\begin{figure}[htbp] 
  \centering
   \includegraphics[width=4in]{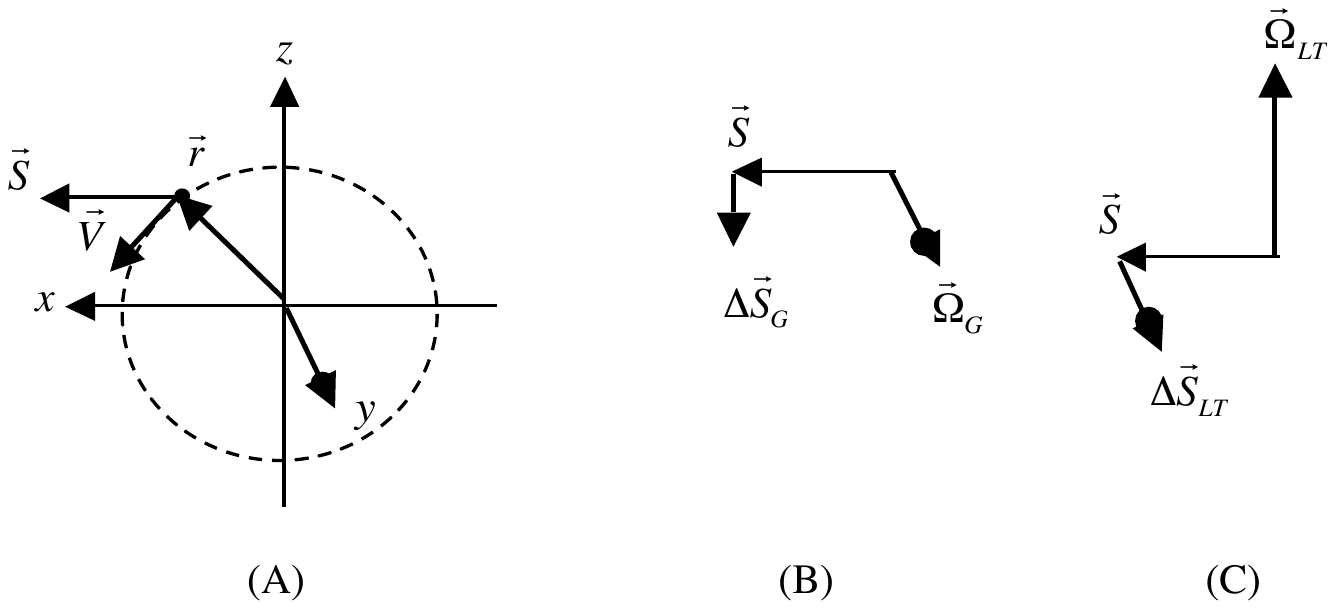} 
   \caption{(A) The orbital and spin orientation vectors. (B) Vectors associated with geodetic precession. (C) Vectors associated with LT precession}
   \label{Fig.1}
\end{figure}


\begin{thebibliography}{99}
\bibitem{17} See chapters 5.1 and 9.6 of reference [10]. Chapter 9.6 contains a sign error that does not affect the answer.
\bibitem{19} See chapter 40.7 of reference [9]. Note that MTW use different PPN parameters, as discussed in reference [6].
\bibitem{43} See chapter 40.7 of reference [9], which uses a different set of PPN parameters.
\end{thebibliography}
\end{document}